%% file: main.tex
\documentclass[acmsmall]{acmart}

\usepackage{graphicx}
\usepackage{graphics}
\usepackage{marginnote}

\AtBeginDocument{%
  \providecommand\BibTeX{{%
    \normalfont B\kern-0.5em{\scshape i\kern-0.25em b}\kern-0.8em\TeX}}}

\acmJournal{PACMHCI}
\acmYear{2022} \acmVolume{6} \acmNumber{CSCW1} \acmArticle{53} \acmMonth{4} \acmPrice{}\acmDOI{10.1145/3512900}

\input{defs}
\usepackage{float}
\usepackage{graphicx}
\usepackage{graphics}

\begin{document}

\title{Characterizing Reddit Participation of Users Who Engage in the QAnon Conspiracy Theories}

\author{Kristen Engel}
\authornote{Both authors contributed equally to this research.}
\authornote{The majority of this work was conducted while the author was at Cornell Tech.}
\affiliation{%
 \institution{University of Washington}
 \city{Seattle, WA}
 \country{USA}
}
\email{engelkri@uw.edu}

\author{Yiqing Hua}
\authornotemark[1]
\affiliation{%
 \institution{Cornell Tech}
 \city{New York, NY}
 \country{USA}
}
\email{yiqing@cs.cornell.edu}

\author{Taixiang Zeng}
\affiliation{%
 \institution{Cornell Tech}
 \country{USA}
}
\email{tz376@cornell.edu}

\author{Mor Naaman}
\affiliation{%
 \institution{Cornell Tech}
 \city{New York, NY}
 \country{USA}
}
\email{mor.naaman@cornell.edu}

\renewcommand{\shortauthors}{Kristen Engel et al.}

\input{00_abstract.tex}




\maketitle

\input{01_intro}

\input{02_relatedwork}
\input{03_datacollect}
\input{04_RQ1}

\input{05_RQ2}
\input{06_RQ3}
\input{07_discussion}
\input{08_conclusion}

\input{98_acknowledgements.tex}

\bibliographystyle{ACM-Reference-Format}
\bibliography{main}

\appendix
\input{99_appendix}

\received{July 2021}
\received[revised]{November 2021}
\received[accepted]{November 2021}

\end{document}

%% file: defs.tex
\usepackage{xspace}

\newcommand{\qusers}{Q-users\xspace}
\newcommand{\uq}{QAnon-enthusiastic users\xspace}
\newcommand{\qauth}{QAnon-interested users\xspace}
\newcommand{\bannedSubs}{QAnon-focused subreddits\xspace}
\newcommand{\bigban}{2018 QAnon ban\xspace}
\newcommand{\activeSubs}{QAnon-enthusiastic active subreddits\xspace}
\newcommand{\preQ}{pre-Q\xspace}
\newcommand{\Q}{Q\xspace}
\newcommand{\postQ}{post-Q\xspace}

%% file: 00_abstract.tex
\begin{abstract}
Widespread conspiracy theories may significantly impact our society.
This paper focuses on the QAnon conspiracy theory,  
a consequential conspiracy theory that started on and disseminated successfully through social media.
Our work characterizes how Reddit users who have participated in \bannedSubs engage in activities on the platform, especially outside their own communities.
Using a large-scale Reddit moderation action against QAnon-related activities in 2018 as  the source, 
we identified 13,000 users active in the early QAnon communities.
We collected the 2.1 million submissions and 10.8 million comments posted by these users across all of Reddit from October 2016 to January 2021.
The majority of these users 
were only active after the emergence of the QAnon conspiracy theory 
and decreased in activity after Reddit’s \bigban. 
A qualitative
analysis of a sample of 915 subreddits where the "QAnon-enthusiastic" users were especially active shows that they participated in a diverse range of subreddits,
often of unrelated topics to QAnon.
However, most of the users' submissions were concentrated in subreddits that have sympathetic attitudes towards the conspiracy theory,
characterized by discussions that were pro-Trump, or emphasized unconstricted behavior
~(often anti-establishment and anti-interventionist).
Further study of a sample of 1,571 of these submissions indicates that most consist of links from low-quality sources,
bringing potential harm to the broader Reddit community.
These results point to the likelihood that the activities of early QAnon users on Reddit were dedicated and committed to the conspiracy, 
providing implications on both platform moderation design and future research.
\end{abstract}

%% file: 01_intro.tex
\section{Introduction}
Widespread conspiracy theories may have detrimental impacts on our society, 
including reducing public trust in science~\cite{douglas2017psychology} and government entities~\cite{kim2016impact},
or radicalizing violent extremist groups~\cite{douglas2019understanding,brotherton2015suspicious}. 
Among them, QAnon emerged as one of the most consequential movements spreading conspiracy theories in the U.S. in recent years.
The QAnon conspiracies are rooted in a narrative suggesting that the world is operated by a group of powerful pedophiles and human traffickers~\cite{nytimesQ}.
Some tenets of this conspiracy are believed by at least 17\% of Americans according to a poll published in December 2020~\cite{nprPoll}.
In June 2021, a terrorism bulletin issued by the U.S. Department of Homeland Security and the Federal Bureau of Investigation listed parts of the QAnon community as domestic violent extremists~\cite{senate}.

Conspiracy theories have attracted an increasing amount of attention from the research community.
Studies have analyzed language patterns of conspiracy-related discussions~\cite{samory2018government,graumann2012changing,phadke2021characterizing},
attributes that correlate with conspiracy group engagement~\cite{phadke2021makes,klein2019pathways},
as well as circumstances that might foster conspiracy beliefs~\cite{mitra2017credibility,stempel2007media,starbird2017examining,koutra2015events,costello2016views}.
In particular,
because of the important consequences that it brings,
recent work has focused on the QAnon movement on fringe social media websites~\cite{papasavva2020raiders}.
However, most of the previous studies focus on the conspiracy theory, or the conspiracy group itself and its activities inside the group. 
An interesting question that was not as well addressed is how members of a conspiracy group 
interact with other online communities that are \textit{not} dedicated to the movement.

In this short paper,
we characterize the broader participation of users who had early involvement with QAnon on Reddit.
Although the conspiracy has spread to many corners of the Internet~\cite{papasavva2021qoincidence},
Reddit was one of the first mainstream social media platforms that had a large QAnon population~\cite{zadroznyQ}.
The social media platform is comprised of subreddits,
communities of users who are interested in particular topics.
In September 2018, Reddit banned a range of QAnon subreddits
for policy violations~\cite{nbc2018ban}. We distinguish these QAnon subreddit bans from others by defining them as the \textit{\bigban}.
Specifically, 
we identified 19 \textit{\bannedSubs} that were shut down by the moderation effort. We use this event to collect a set of users who were active in these banned subreddits.
We define the 13,182
users who have posted submissions in these 19 \bannedSubs as \textit{\qusers}.
We further describe two important subsets of the \qusers. We define \textit{\uq} as the top 25\% most active \qusers;
authors who have published five or more submissions in the 19 \bannedSubs.
There are 3,506 such \uq. We refer to the remaining 9,676 users who do not meet the threshold as \textit{\qauth}.
We collected all the 2.1 million submissions and 10.8 million comments
of all \qusers on Reddit, from October 28th, 2016, one year prior to the first Q drop \cite{wangBI, wiki:QAnon},
to January 23rd, 2021, 72 hours after the U.S. Presidential Inauguration.  

With this data, we first characterize 
activity patterns of \qusers in relation to the emergence of the QAnon conspiracy and to the
\bigban.
We show that the majority of these users
were significantly more active on Reddit during the period between the emergence of the QAnon conspiracy and the
\bigban, 
as compared to the period before the emergence, \textit{and} the period after the ban.
Chandrasekharan et. al.~\cite{chandrasekharan2017you} have shown that community bans are effective on Reddit in terms of reducing platform participation of users from the banned subreddits.
Our findings validate this result, and indicate that the tendency of \qusers' participation, 
at least in the early period of the movement, 
was driven by their interest in the QAnon conspiracy. 

Looking more generally at the activity of \qusers beyond their community,
we analyzed the other subreddits where they were active.
To this end, we sampled 915 out of the 12,987
subreddits where  
at least two \uq 
have authored submissions. 
We categorized the sampled subreddits according to the topics that they focus on
and the relation of these subreddits to the QAnon conspiracy.
Our results indicate that the 
majority of subreddits that \qusers concentrate activity in
focus on content sympathetic towards the QAnon conspiracy,
such as pro-Trump rhetoric or centering
unconstricted behavior, such as hate speech, pro-freedom of speech, and anti-establishment discourse.
In particular, our results show that more activity had shifted to these types of subreddits after the \bigban.

Finally, we characterize the type of participation and harmful content posted by \qusers across different communities.
We qualitatively analyzed 1,571 submissions from 
these users, sampled from the categorized subreddits.
Our findings suggest that 
\qusers might have influence on other subreddits, by acting as moderators, 
sometimes in subreddits that are of topics that are misaligned with the QAnon conspiracy.
The majority of the harmful content 
these users post across communities contain links to low-quality sources.
Surprisingly, very few of the submissions that we sampled were characterized as negative interactions and hate speech, even when \qusers post in subreddits where users are likely to hold opposite views.

Our findings provide better understanding on the consequential QAnon conspiracy movement,
as well as on how the early, dedicated members of that conspiracy community interact with a broader audience.
The discoveries may also help inform platform moderators to take stronger actions to prevent further harms.

%% file: 02_relatedwork.tex
\section{Background and Related Work}

The QAnon conspiracy has received significant recent research attention, including its activity on Reddit and other social media.
We provide some relevant background on the QAnon conspiracy community and its relation with Reddit.
We then review related work that examine conspiracy theories, in particular QAnon's activities on social media,
as well as work on the efficacy of deplatforming deviant communities.

\paragraph{The QAnon conspiracy theory and Reddit.}
The QAnon conspiracy theory refers to a series of narratives that emerged in October 2017, when an anonymous poster known as "Q" posted what became known as "Q drops"~\cite{wangBI,wiki:QAnon} on the fringe image-board website 4chan~\cite{bernstein20114chan}.
The core of the conspiracy theory asserts that former U.S. President Trump 
would use his executive powers to arrest
a cabal comprised of pedophiliac Satan-worshiping elites from Hollywood to the U.S. Government~\cite{nytimesQ}.
Since its inception, despite failed prophecies and a lack of supporting evidence, the QAnon conspiracy has evolved to incorporate a more expansive and sometimes conflicting set of conspiracies. These theories center around a supposed "deep state" and a "global elite" and range from predictions, such as the re-emergence of a deceased John F. Kennedy Jr., to responses to current events, such as claims of election fraud and the "plandemic" \cite{yahooJFK, abcElectFraud, bbcQcovid}.
According to the Pew Research Center~\cite{pewQ}, by September 2020, 47\% of the Americans have heard of the conspiracy theory.
A separate survey conducted in December 2020 shows that 37\% of the Americans were unsure whether QAnon was true or false while 17\% believed in some of its tenets~\cite{nprPoll}.
QAnon adherents are feared 
to threaten and incite violence~\cite{nbc2018ban, mob2019killer, hoover2020block, fbiViolent}
and are listed by U.S. law enforcement to be tied to domestic violent extremism~\cite{senate}.

After QAnon's success in propagating on fringe platforms, 
a community was created on Reddit in December 2017 to focus on QAnon discussions~\cite{cbtsstream}.
Reddit, a mainstream social media platform, consists of over two-million 
topic-focused content generating and sharing communities, known as subreddits.
Users can make submissions, or comment on other users' submissions in subreddits.
On September 12, 2018,
Reddit banned several QAnon subreddit communities in response to repeated violations of the site's Content Policy~\cite{nbc2018ban}.
Reporting suggested that the \bigban was successful in terms of reducing QAnon narratives on Reddit~\cite{atlanticQ}.

\paragraph{Conspiracy communities on social media.}

Social media clearly contributes to the accelerated propagation of conspiracy content. 
The algorithms create echo-chambers that promote conspiracy theories~\cite{faddoul2020longitudinal}.
Platforms foster trust within social circles~\cite{friggeri2014rumor,allport1947psychology} that spread rumors,
and support the human nature to bond through collective sense-making and gossip~\cite{kou2017conspiracy}.
Prior work examining conspiracy communities on Reddit have focused on language patterns~\cite{samory2018government,shahsavari2020conspiracy, phadke2021characterizing}
and user engagement inside the conspiracy communities~\cite{samory2018conspiracies}.
Most related to our research, Phadke et. al.~\cite{phadke2021makes} studied factors that might draw Reddit users into conspiracy groups,
including interacting with current members of the conspiracy group.
While their work focuses on predictors for engagement within conspiracy communities, our study characterizes the engagement of a conspiracy community across Reddit.

Given its importance, a growing body of research focuses on the QAnon conspiracy theory, 
including characterizing the QAnon movement through analysis of Q drops and QAnon user social imaginaries~\cite{aliapoulios2021gospel,phadke2021characterizing}.
Researchers have explored QAnon users within fringe~\cite{papasavva2021qoincidence}
and instant messaging-based social media~\cite{hoseini2021globalization},
and examined the cross-platform movement from fringe to mainstream social media~\cite{de2020tracing}.
Previous work has also pointed out that 
social media platforms not only contribute to disseminating QAnon content that incites or leads to violence ~\cite{amarasingam2020qanon}
, but also support the movement's collective sense-making~\cite{partin2020construction}.
Different from prior work,
we focus on QAnon users' behaviors on a mainstream platform interacting with a diverse range of communities,
and on how these users were impacted by the platform's moderation action.

\paragraph{The effectiveness of deplatforming.} 
Prior work has shown that for social media websites,
deplatforming is a powerful tool to reduce harmful behaviors on the platform~\cite{jhaver2021evaluating,chandrasekharan2017you,saleem2018aftermath}. 
However, users or communities that were deplatformed might move to alternative platforms~\cite{rogers2020deplatforming,rauchfleisch2021deplatforming} 
and engage in more toxic behaviors~\cite{ali2021understanding,ribeiro2020does}.
Nevertheless,
on Reddit, researchers have examined platform-level actions such as subreddit quarantines~\cite{chandrasekharan2020quarantined, shen2019discourse} and subreddit bans~\cite{chandrasekharan2017you, newell2016user,saleem2018aftermath},
showing the effectiveness of both types of actions within the platform.
Different from previous studies, 
we characterize QAnon users' participation on Reddit in other communities 
from a period before the creation of the \bannedSubs and through a period after the \bigban.
Further, 
our study is not only focused on deplatforming effectiveness, as we are interested more generally in the activities of 
\qusers in other Reddit communities.

%% file: 03_datacollect.tex
\section{Data Collection}
To better understand QAnon participants' activity patterns on Reddit, 
we collect a multi-part dataset. 
The core of our dataset consists of 2,099,875
submissions and 10,831,922
comments from 63,697 subreddits, posted by 13,182 \qusers identified from 19 banned \bannedSubs.
This set of subreddits was not made public, but we explain how we extracted it below.

To find \qusers, we first had to identify the \bannedSubs.
On September 12, 2018, 
Reddit coordinated a large-scale intervention against \bannedSubs for platform policy violations~\cite{nbc2018ban}.
To list these impacted subreddits, we start from a large such community, which had been identified in media reports~\cite{nbc2018ban} as one of the banned communities: r/greatawakening. 
At the time of the \bigban, r/greatawakening had 12,862 submission authors and over 71,000 subscribers~\cite{nbc2018ban}. 
We use r/greatawakening as a seed community and search for subreddits that are of a similar user base.
We search for all subreddits satisfying the following three conditions:
\begin{itemize}
\item The subreddit shares more than one submission author with r/greatawakening. 
\item 
The subreddit name contains at least one of the QAnon-associated terms\footnote{QAnon-associated terms: qproofs, calmbeforethestorm, qanon, greatawakening, stqrm, cbts, wwg, wga, cabal, outoftheshadows, enjoytheshow, awake, thestorm, q, theshow}, which we 
identified from hashtags on Reddit, Twitter, Facebook, and Instagram,
converted to lowercase, and lemmatized.
\item The subreddit contains at least one submission within 45 days prior to the \bigban and no further activity after. 
\end{itemize}
In total, we identify 19 \bannedSubs, including r/greatawakening. 
The list of subreddits, number of submission authors,
submissions, and reasons for the ban for 
each subreddit are listed in Table~\ref{tab:bannedSubs} in Appendix~\ref{ap:codebooks}.

Most of these discovered subreddits were small in size, with r/greatawakening containing over 18 times as many users as of the next largest community.
The remaining 18 \bannedSubs 
have 78 submission authors on average. Using this set of 19 subreddits,
we identify a total of 13,182 \qusers who have made at least one submission in any of the subreddits.

Given this initial set of users, we expand the dataset to include the \qusers' activities in other subreddits.
We use the Pushshift API~\cite{baumgartner2020pushshift, psawPython} to collect all the comments and submissions
of \qusers during a 50-month period.
The data collection period starts on October 28th, 2016, one year prior to the first Q drop \cite{wangBI, wiki:QAnon},
and ends on January 23rd, 2021, 72 hours after the U.S. Presidential Inauguration.
The 2021 inauguration was chosen as it was a significant event for the QAnon conspiracy movement, and marked the end of a period with potentially heightened activity and attention.
In total, our dataset includes 2,099,875
submissions and 10,831,922 comments from 
63,697 subreddits.
We refer to the subset of 12,987 subreddits where at least two of the \uq have made a submission as the \textit{\activeSubs}, which include 1,554,183 submissions and 9,687,789 comments made by \qusers.

In summary, our extensive dataset includes users who were active in \bannedSubs banned by Reddit, as well as the activity of these users in other communities for a 
period starting before the emergence of the QAnon conspiracy and continuing for years after the \bigban. The data are described and available on Github\footnote{https://github.com/sTechLab/QAnon\_users\_on\_Reddit}.

\paragraph{Ethical considerations.}
Our work uses data in the public domain, collected through a publicly available APIs from Reddit and Pushshift. Further, the Pushshift API provides Reddit users with a mechanism to request to remove their account and its content from the dataset.
We recognize that some users may not be aware that their data is accessible and used in this way, so we hashed usernames to further obscure identification. 
Additionally, we focused our analysis on the aggregation of the user base. 

%% file: 04_RQ1.tex
\section{Activity patterns of \qusers}
\label{sec:activity}

We first study the activity patterns of \qusers across all Reddit communities, and how these patterns changed 
with the emergence of QAnon conspiracy theory and the \bigban. 
For \qusers, 
the QAnon conspiracy theory may have motivated their participation on Reddit.
On the other hand, 
moderation actions~\cite{chandrasekharan2017you,saleem2018aftermath} against them, 
such as the \bigban, may have a negative impact on their future participation.
We analyze the basic account statuses and activity patterns of \qusers with unrestricted accounts, or existing accounts without limitations on participation.
We obtain user account
status~(unrestricted, suspended, and does not exist)
for all \qusers as of June 2021.
We find that the majority of 
\qusers were significantly less active before the QAnon conspiracy theory emerged, 
indicating the possibility that the motivation of their participation on Reddit was heavily influenced by the conspiracy.
Further, many of the \qusers, even those who stayed on the platform, decreased activity by a large extent after the 2018 QAnon ban by Reddit,
likely due to the effect of moderation.

We refer to the \textit{user's activity} as the total number of submissions and comments by a Q-user
in any subreddit.
The \textit{\Q period} spans the 319 days between the first "Q drop" on 4chan, 
on October 28th, 2017 
and the moderation action from Reddit to ban \bannedSubs
on September 12, 2018.
For ease of comparison, we use same-length before-and-after time periods for our analysis here. 
The \textit{\preQ period} spans the 319 days right before the \Q period, beginning on 
December 13, 2016. The \textit{\postQ period} spans the 319 days right after the \Q period, ending on July 28, 2019.

\begin{table}[t]
\footnotesize
    \caption{\qusers on Reddit.}
  \begin{tabular}{lr|rrr|rr}
\toprule                &  Total &  \multicolumn{3}{|c|}{Account Status~(as of June, 2021) }& \multicolumn{2}{c}{Among unrestricted users} \\
                &   &  Unrestricted &  Does Not Exist &  Suspended & Active: Pre-Q &  Active: Post-Q \\ 
\midrule
   QAnon-interested &   9676 &    8081 (83.52\%) &            1156 (11.95\%) &        439 (4.54\%) &            2702 (33.44\%) &            4611 (57.06\%) \\
 QAnon-enthusiastic &   3506 &    2933 (83.66\%) &             468 (13.35\%) &        105 (2.99\%) &             855 (29.15\%) &            1951 (66.52\%) \\
\bottomrule
\end{tabular}
\label{tab:userStats}
\end{table}

Table~\ref{tab:userStats} shows the total number of \uq and \qauth, as well as a June 2021 snapshot of the account status of the users in each group: the number of users from each group that are unrestricted, that do not have an existing account~(indicates account deletion), or that are suspended~(indicates a moderation intervention of account suspension or ban).
Note that more of the \qauth were suspended~(4.54\%)  compared to the \uq~(2.99\%).
The table indicates that among all the users who have unrestricted account status,
majority of the users from both groups were not active before the emergence of Q~(66.56\%
of the unrestricted \qauth and 
70.85\%
of the unrestricted \uq). After the \bigban, around one-third of these users disengaged with Reddit.
For example, among the unrestricted \uq, 66.52\%
of them remained active on Reddit.

\begin{figure}[b!]
\centering
    \includegraphics[width=\linewidth]{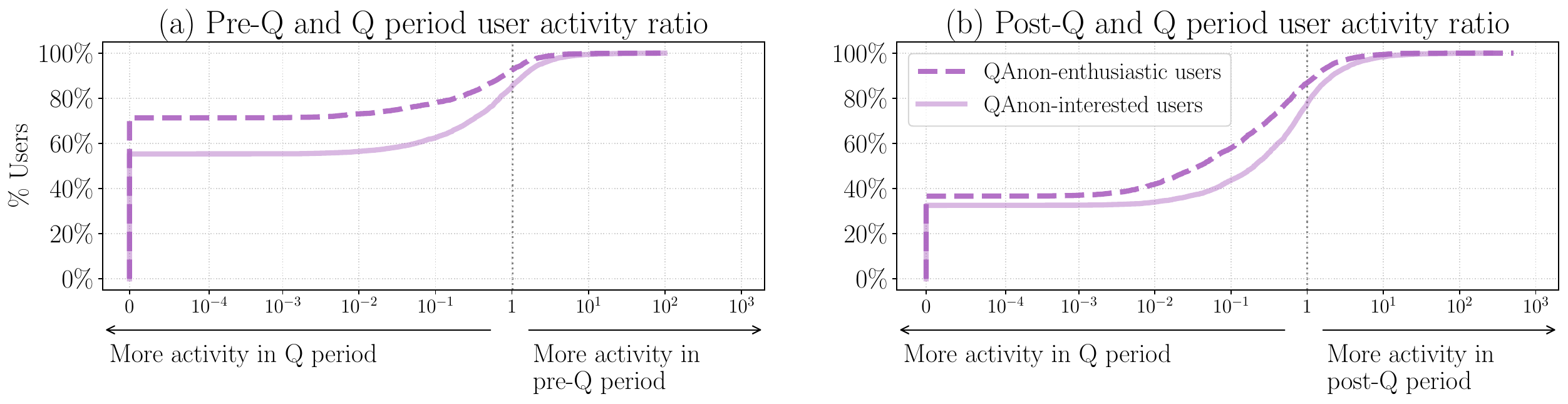}
    \caption{\qusers are more active in the \Q period. (a) The ratio between a user's activity in the \preQ period over that in the \Q period. (b) The ratio between a user's activity in the \postQ period over that in the \Q period.}
    \label{fig:activity}
\end{figure}

While some \qusers were on Reddit before and after the \Q period, we set
to understand how \qusers' activity patterns differed before and after that period. 
We compare these users' activity in three consecutive time periods 
of equal length
, the \preQ, \Q, and \postQ periods.
Figure~\ref{fig:activity} shows a comparison of the relative activity of each user in the \preQ period (a) and \postQ period (b) with that in the \Q period.
We restrict users to those who posted at least 5 submissions or comments in the \Q period across all of Reddit, and whose account statuses were unrestricted at the time of our data collection, which consists of 58\% of the \qauth~and all of the \uq.
We use the ratio of activity between pre- (or post-) and the \Q period; when users are more active in the \Q period, that ratio will be smaller than $1$ (on the left side of each figure).
The X-axis represents a range of user activity ratio, from $10^{-4}$ to $10^3$ 
and the Y-axis represents the cumulative percentage of \qusers with a given activity ratio.
Most \qusers were significantly more active in the \Q period.
For example, looking at the \preQ figure, and the X-axis at a user activity ratio of $10^{-1}$, we see that 63\% and 78\% of the \qauth and \uq~(respectively)
were at least ten times more active in the \Q period as compared to in the \preQ period.
Similarly, in the \postQ figure, 44\% and 58\% of the \qauth and \uq~ 
(at a user activity ratio of $10^{-1}$)
were at least ten times more active in the \Q period as compared to in the \postQ period. 
This result echos our observation in Table~\ref{tab:userStats}, showing that many of the \qusers increased activity after the emergence of the QAnon conspiracy theory. 
The results also show that even if \qusers were still active on the platform, their activity decreased across Reddit after the \bigban.

%% file: 05_RQ2.tex
\section{Characterizing Communities with Q-Users' Submissions}
\label{sec:sample}

Here we focus our analysis on \textit{where} the 
\qusers were active on Reddit beyond the \bannedSubs throughout the 50-month study period. 
We use qualitative analysis to characterize the \activeSubs, where at least two \uq have published submissions.
Our analysis is centered on \uq, the most active \qusers, who contributed five or more submissions as defined above.
Our results show that although the majority of the sampled \activeSubs are of content unrelated to QAnon, the majority of 
activities of these users
are concentrated in the communities that are sympathetic to
or aligned with the QAnon narrative.
In particular, their submissions are in the communities that cover topics that are supportive of Trump or about unconstricted behavior, such as hate speech, pro-freedom of speech, and anti-establishment.

The submissions in \activeSubs accounted for over 
89\% of all submissions by \uq on Reddit during that period. 
In total, this activity amounts to 923,942 submissions by 2,636 \uq in \activeSubs during the study period.
To understand the activities of these most active users 
we sample and categorize 915 \activeSubs according to their topic,
and the relation between the sampled subreddits with the QAnon conspiracy theory. 
The sample consists of communities with varying sizes and different estimated percentage of \uq in the community.

\paragraph{Methods.} 
We begin by sampling \activeSubs for 
qualitative analysis.
Our sampling strategy ensures that we cover subreddits with different types of engagement from the general public as well as that from \uq.
First, to ensure coverage of the subreddits where \uq are most active, 
we sample the top 3\% of the \activeSubs with the most \uq' submissions.
This set includes 390 subreddits,
each with at least 125 submissions from \uq.
Second, 
we group the remaining \activeSubs into five stratified bins,
according to the number of subscribers that they have. We obtain subscriber counts using the Reddit API~\cite{prawPython} as of June 2021. For private, quarantined, or banned subreddits, we use Metrics For Reddit~\cite{metrics2021reddit} to retrieve the number of subscribers. For 88 (0.68\% of all) subreddits, none of these were available,
we use the median of all known subscriber counts of the \activeSubs as an estimate.
To estimate the levels of \uq infiltration in subreddits,
we compute the number of the subreddit's \uq
submission authors 
over the number of the subreddit's subscribers.
Inside each of the five bins, we group the subreddits into three strata according to this ratio.
To create our sample of subreddits, we randomly sample 35 subreddits from each of the fifteen ($5 \times 3$) groups, 
yielding a set of 525 subreddits. Finally, we combine the set of subreddits from the top 3\% and the set from the stratified random sampling to 
obtain a sample of 915 \activeSubs, which includes 85\% of the \uq' submissions and 58\% of all \qusers' submissions in all \activeSubs.

We use an inductive thematic analysis~\cite{braun2006using} to understand the different 
content focuses and types of the subreddits in our sample. 
The analysis is based on key information of each subreddit: the subreddit name, public description, and the most recent top five upvoted "hot" designated submissions.
Two authors independently reviewed the data and conducted four rounds of iterative coding and reconciliation to produce a subreddit codebook (Appendix~\ref{ap:codebooks}, Table~\ref{tab:subCodes}).
Two authors coded the sampled subreddits, each with one or more of the 24 codes. The Cohen’s kappa for the this coding of the subreddits is 0.83. 
After this step, the two authors resolved conflicts in the code through discussion.
We grouped the 
24 codes in the book in two different dimensions: (1) nine \textit{topic labels} derived from common subreddit themes; and (2) six \textit{relation labels} representing the relationship of the subreddit with the QAnon narrative.
The mapping of codes to relation and topic labels are detailed in the codebook (Appendix~\ref{ap:codebooks}, Table~\ref{tab:subRelation} and  Table~\ref{tab:subTopics}, respectively). 
For example, subreddits with the 
codes
\textit{conspiracy - other} and \textit{politics - right} are assigned a relation label of \textit{Sympathetic} due to having similar views or supporting common QAnon themes. 
In addition to the relation label, subreddits with the 
code
\textit{conspiracy - other} were assigned a topic label of \textit{Unseen/unknown}, capturing subreddits where the community speculates or searches for meaning and explanation of different phenomena (see Appendix~\ref{ap:codebooks}, Table~\ref{tab:subTopics}).
The code \textit{politics - right}, to use another example, was assigned a topic label of \textit{Pro-Trump} in addition to its relation label.

\begin{figure}[t!]
\centering
    \includegraphics[width=\linewidth]{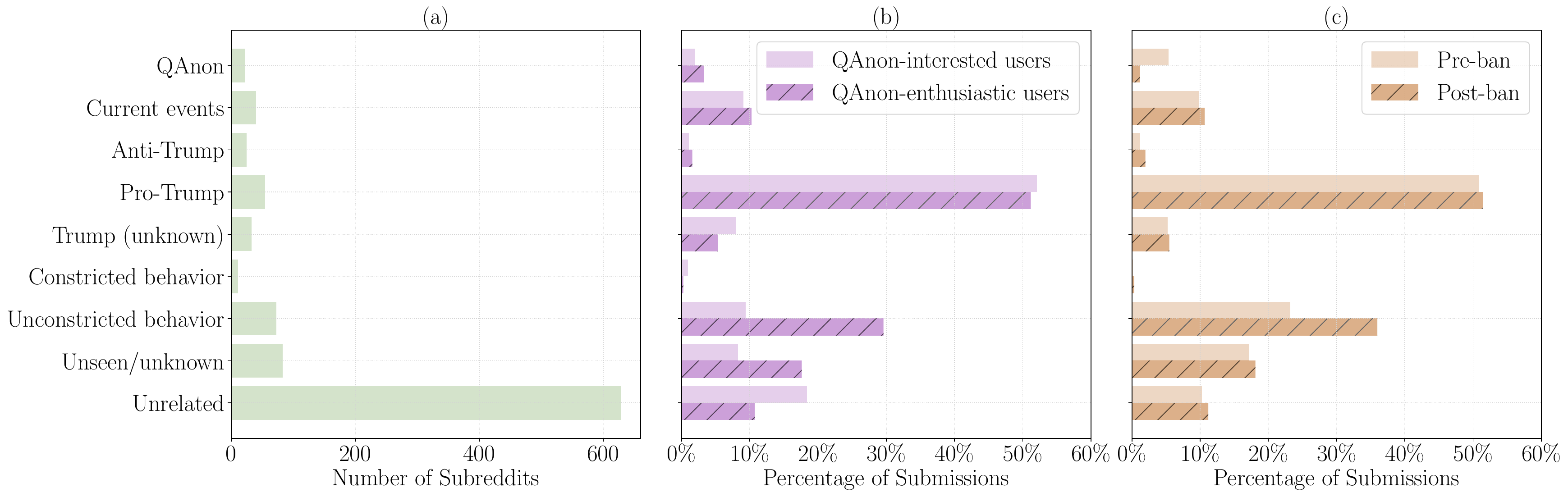}
    \caption{Sampled \activeSubs with different topic labels: (a) Number of subreddits with content of different topics. (b) \uq' and \qauth' submissions in subreddits with different topics. (c) \uq' submissions in subreddits with different topics before and after the \bigban.}
    \label{fig:subreddit_topic}
\end{figure}

\paragraph{Topic labels.}
Figure~\ref{fig:subreddit_topic} shows the analysis of sampled \activeSubs according to their topic label. 
Figure~\ref{fig:subreddit_topic}(a) shows the count of subreddits in our sample where \uq were active during the entire 50-month study period,
categorized by the topic labels.
The X-axis represents the number of subreddits that contain the type of topic specified by the Y-axis.
The figure shows that \uq were broadly active: most of the sampled subreddits~(629 out of 915) contain \textit{Unrelated} topics~(bottom row).

While \uq were active in many different types of subreddits, most of them of \textit{Unrelated} topics, the majority of their \textit{activity} was focused on a number of subreddits with topics relevant to QAnon.
In total, \qauth and \uq posted 421,908
and 794,423 submissions in all sampled subreddits respectively.
Figure~\ref{fig:subreddit_topic}(b) shows the percentage of submissions from \qauth and \uq in subreddits with different types of topics.
The X-axis represents the percentage of total submissions posted by each user group in a subreddit with the type of content specified by the Y-axis.
For example, the bottom row of Figure~\ref{fig:subreddit_topic}(b) shows that only 18\% of the \qauth' submissions~(bar with lighter shade) and 11\% of the \uq' submissions~(bar with darker shade and stripes)
were posted in subreddits that are about topics unrelated to QAnon. 
Most of the submissions from both groups are concentrated in subreddits with \textit{Pro-Trump} topics~(52\% for \qauth and 51\% of \uq).
Compared with \qauth,
more of the \uq' submissions are in subreddits with topics of \textit{Unconstricted behavior}~(29\% vs. 9\%).
The difference is significant~($\chi^2(8, N=1,491,678)=82,077.3, p<0.001$).

Figure~\ref{fig:subreddit_topic}(c) shows the percentage of 
\uq' submissions before and after the \bigban 
over the entire study period, in subreddits with different types of topics.
\uq posted 397,752 and 396,671
submissions in these sampled subreddits
before and after the \bigban.
The X-axis represents the percentage of total submissions from one of the time periods specified by the legend.
For example, the fourth row from top shows that 51\% of the 
\uq' submissions before the ban~(bar with lighter shade) and 52\% of the 
\uq' submissions after the ban~(bar with darker shade and stripes
posted in subreddits that focus on \textit{Pro-Trump} topics.
Similarly, there is increased activity from \uq after the ban on submissions in subreddits we labeled as  \textit{Unconstricted behavior}.
The difference is significant~($\chi^2(8, N=1,029,581)=20,398.2, p<0.001$).

In conclusion, our results indicate that although most of the \activeSubs 
are of unrelated topics,
the majority of the \uq' submissions are concentrated in subreddits with \textit{Pro-Trump} 
and \textit{Unconstricted behavior} content.
After the \bigban, \uq' submissions in \textit{Unconstricted behavior}-focused subreddits increased significantly
while those in \textit{Pro-Trump}-focused subreddits 
increased only slightly.
One possible explanation for the shift being relatively minor could be the banning in June 2020 of the largest \textit{Pro-Trump} subreddit, r/The\_Donald, where 41.5\% of \uq authored submissions~\cite{thedonald2020wsj}.

\begin{figure}[t!]
\centering
    \includegraphics[width=\linewidth]{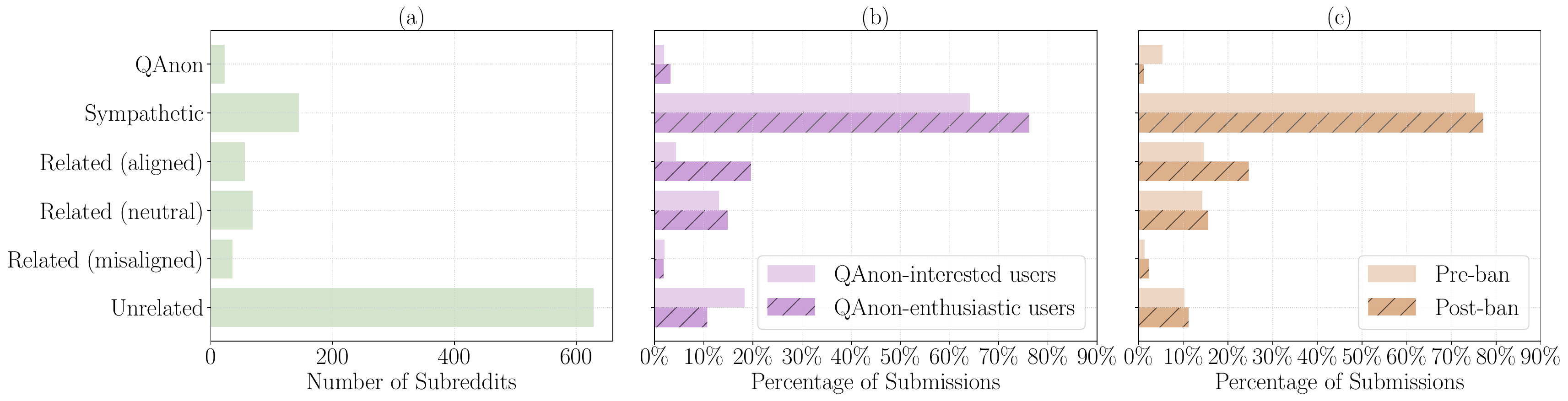}
    \caption{(a) Sampled \activeSubs with different relation to QAnon. (b) \uq' and \qauth' submissions in subreddits with different relation to QAnon. (c) \uq' submissions in subreddits with different relations to QAnon before and after the \bigban.}
    \label{fig:subreddit_relation}
\end{figure}

\paragraph{Relation labels.} 
Figure~\ref{fig:subreddit_relation} shows the analysis of where \uq were active in the sample according to the subreddits' relation labels, a direct evaluation of how the subreddit relates to the QAnon narrative. 
Figure~\ref{fig:subreddit_relation}(a) shows the number of subreddits with different types of relation labels.
The X-axis represents the number of subreddits in our sample for each type of relation, specified by the Y-axis.
Similarly, most of the sampled subreddits~(629 out of 915) 
are of \textit{Unrelated} relation~(bottom row). 

Figure~\ref{fig:subreddit_relation}(b) shows the percentage of submissions from the two user groups, \qauth and \uq, in subreddits with different types of relations to the QAnon narrative.
The X-axis represents the percentage of total submissions from one of the user group specified by the legend, that were posted in a subreddit with the type of relation
specified by the Y-axis.
Similar to our previous findings, most of the submissions from both groups are concentrated in subreddits that are \textit{Sympathetic} towards QAnon~(64\% for \qauth and 76\% of \uq).
Further, compared with \qauth,
more of the 
\uq' submissions are in subreddits with QAnon-related topics that are \textit{Related (aligned)} with the narrative~(19\% vs. 4\%).
The difference is significant~($\chi^2(5, N=1,444,019)=56,617.7, p<0.001$).

Figure~\ref{fig:subreddit_relation}(c) shows the percentage of 
QAnon-enthusiastic users' submissions before and after the \bigban, in subreddits with different types of relations.
The X-axis represents the percentage of total submissions from one of the time periods specified by the legend, that were posted in a subreddit with the type of relation
specified by the Y-axis.
For example, the 
second row from top shows that 75\% of the 
\uq' submissions before the ban~(bar with lighter shade) and 77\% of the 
\uq' submissions after the ban~(bar with darker shade and stripes)
were posted in \textit{Sympathetic} subreddits.
There's also an increase in the portion of the \uq'~submissions after the ban in subreddits that are \textit{Related (aligned)} to the QAnon narrative. 
The difference is significant~($\chi^2(5, N=1,003,887)=20,575.7, p<0.001$).

Our findings show that the majority of \uq' submissions are concentrated in subreddits that are \textit{Sympathetic}
towards the QAnon conspiracy theory,
especially when compared with \qauth.
After the \bigban, \uq concentrated more in subreddits that 
share similar elements, views, or themes with QAnon, that were not directly misaligned with or focused on the conspiracy. 

%% file: 06_RQ3.tex
\section{\qusers' Activities Across Communities}

\qusers interact with different types of subreddits in different ways. 
One of the more concerning types of interactions is when these users serve as moderators of subreddits, where they can exert outsize influence on these communities. 
Another interaction that may be concerning is the posting of content that is harmful within these subreddits. 
In this section, we explore these two types of interactions.
Our analysis shows that a non-trivial number of subreddits could potentially be heavily influenced by \qusers who were serving as their moderators.
This phenomenon is not limited to subreddits that are of
similar views or themes to QAnon, but is also extended to subreddits that promote opposing or disjointed content.
Our results below also show that \qusers spread low-quality information across all other subreddit types, although we detected 
less
such activity in subreddits that are \textit{Unrelated} to QAnon.

\paragraph{\qusers as moderators.}
We examine the different types and characteristics of
the \activeSubs where \qusers act as moderators. We obtain the moderator names as of June 2021 using the Reddit API~\cite{prawPython}.
In total, out of 12,987 \activeSubs, we found that 88 \qauth and 143 \uq are moderators of 113 and 371 subreddits, nine of which have over 1,000,000 subscribers including r/worldnews and r/politics. 
In particular, 240 of these subreddits have moderation teams solely comprised of \qusers.

\begin{figure}[t!]
\centering
    \includegraphics[width=\linewidth]{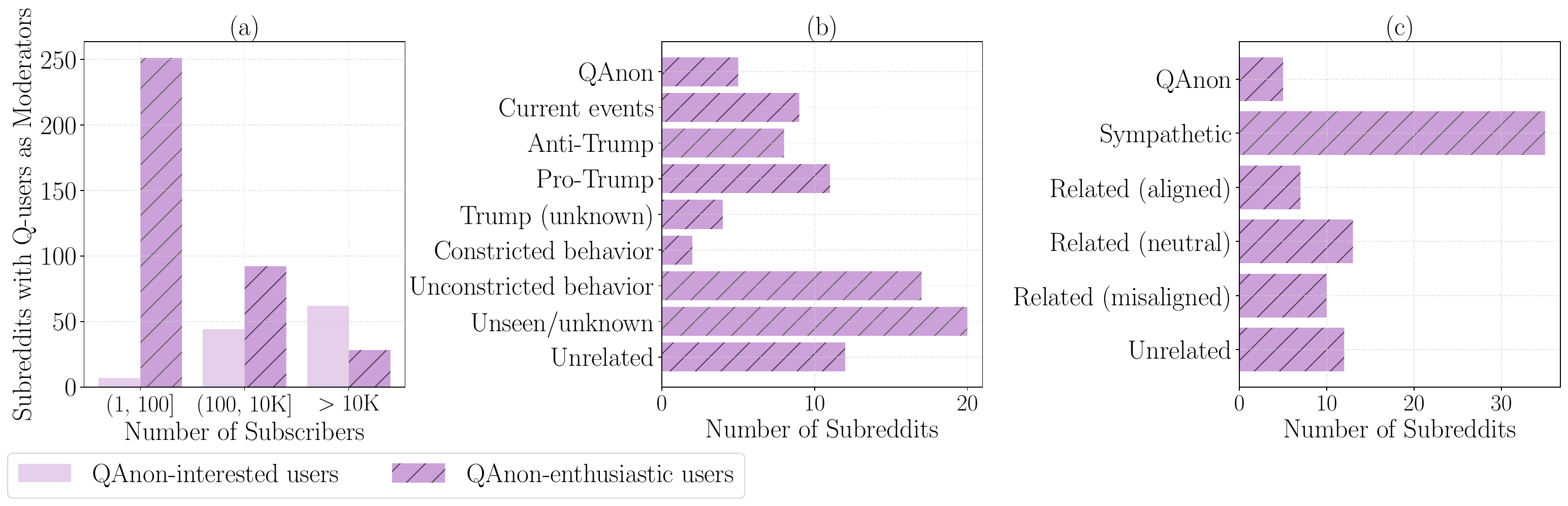}
    \caption{(a) Number of \activeSubs with \qusers as moderators, with varying number of subscribers. Number of sampled \activeSubs that have \uq as moderators, (b) with contents of different topic labels and (c) of different relation labels. }
        \label{fig:subreddit_mod}
\end{figure}

Figure~\ref{fig:subreddit_mod}(a) shows the number of \activeSubs with at least one Q-user
as moderator
(Y-axis) by number of subscribers that these subreddits have (X-axis).
The legend separates \qauth~(bar with the lighter shade)
and \uq~(bar with the darker shade and stripes.)
For example, the first
column from left shows that most of the \activeSubs~(251) that 
have \uq as moderators are small communities, with fewer than 100 subscribers. 
At the same time, 
several larger communities 
with more than 10,000 subscribers have moderators who are \qauth~(62 subreddits, first bar with the lighter shade from right) or \uq~(28 subreddits, first bar with the darker shade and stripes from right).
These numbers indicate that both \uq and \qauth might have large impact on the broader community and conversation on Reddit. 

To understand the types of 
subreddits where these users serve as moderators, we examine our sample of
915 \activeSubs~(as described as in Section~\ref{sec:sample}).
Within this sample, we find that~38 \uq are moderators of 69 of these subreddits, and~31 \qauth serve as moderators of 29 of the 915. 
Furthermore, 25 of the 915 sampled subreddits have moderation teams solely comprised of \qusers.
Figure~\ref{fig:subreddit_mod}(b) and (c) break down the types of subreddits moderated by these users according to the topic labels and relation labels (Y-axis) respectively. 
The X-axis represents the  
number of subreddits moderated by
\uq for each type of relation and topic label.
For example, in Figure~\ref{fig:subreddit_mod}(c), the majority of the subreddits with \uq as moderators are 
\textit{Sympathetic} to the QAnon conspiracy theory~(35 subreddits, second column from top.)
However, as shown in Figure~\ref{fig:subreddit_mod}(b),
there are cases where \uq moderate subreddits with topics that are not clearly aligned with the QAnon narratives,
such as \textit{Current events}~(9 subreddits), \textit{Unseen/unknown} 
(20 subreddits),
or even subreddits with topics that are clearly misaligned with the QAnon narratives,
such as \textit{Anti-Trump} 
(8 subreddits).

\paragraph{Content Posted by \qusers.} 
In total, 
the \qusers made 1,554,183 submissions in the 12,987 \activeSubs over the 50-month study period.
To better understand what type of harmful content \qusers post 
in other subreddits than the QAnon-focused ones,
especially inside the \activeSubs, 
we apply a labeling scheme to a sample of 1,571 
Q-user-authored submissions.
These submissions are extracted from the 915 sampled \activeSubs.

We used a stratified random sampling strategy to obtain the submissions sample in order to ensure a balance across subreddit relation labels and subreddit topics.
For each relation label $R$ and topic label $T$ pair, we sample a number of submissions is proportional to the number of subreddits labeled with both $R$ and $T$. 
In total, we sample 1,571 submissions made by \qusers in the sample of 915 \activeSubs.

\begin{figure}[t!]
\centering
    \includegraphics[width=\linewidth]{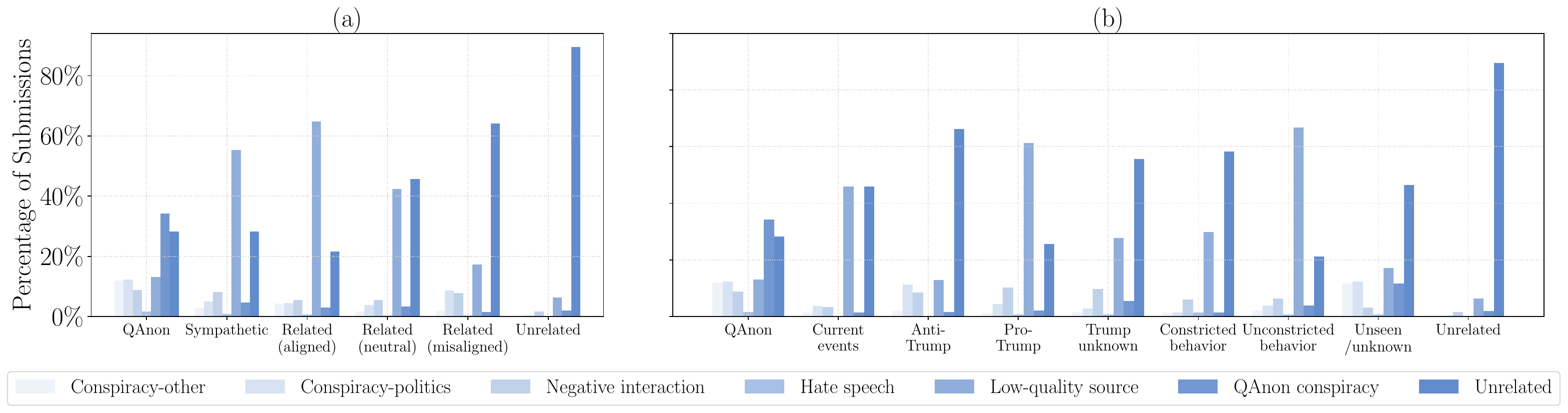}
    \caption{Types of harmful submissions (different color bars) in sampled \activeSubs according to their frequency in subreddits of (a) different topic labels, and (b) different relation labels.}
    \label{fig:submission}
\end{figure}

The 1,571 sampled submissions were assigned one or more labels indicating if the post contained different types of harmful content: \textit{QAnon conspiracy}, \textit{Conspiracy-politics}, \textit{Conspiracy-other}, \textit{Hate speech}, \textit{Negative interaction}, and \textit{Low-quality source}. We expand on these labels in Appendix~\ref{ap:codebooks}, Table~\ref{tab:submission_code}. 
Two authors independently reviewed and labeled the submissions.
If a post did not contain one of these pre-defined harms, it was labeled \textit{Unrelated}. 
To apply the \textit{Low-quality source} label we use Media Bias Fact Check~\cite{mbfc}~(MBFC, accessed June 2021),
a resource frequently used in previous work~\cite{patricia2019link, singh2020understanding, zhou2020recovery}. The authors labeled a link submission as a \textit{Low-quality source} if it was labeled by MBFC as low credibility, mixed, low or very low factual reporting, questionable sources, or conspiracy-psuedoscience. In total, 55\% of the unique links in our sample were included in MBFC. 
The Cohen's kappa for labeling was 0.53, followed by a reconciliation process that produced the final labels used in the analysis.

To understand where \qusers promoted harmful content, we examine these labeled submissions across different types of subreddit.
Figure~\ref{fig:submission} shows the different types of submissions posted by \qusers
in subreddits of (a) different topic labels and (b) different relation labels.
The Y-axis represents the percentage of the different types of submissions. The submission types are the columns, using different shades of blue for each type as specified in the legend. 
The percentage is the ratio of this type of submission out of all the sampled submissions for that category of subreddits.
For example, 
in subreddits
\textit{Related (neutral)}, \textit{Related (misaligned)}, and \textit{Unrelated} to
QAnon narratives,
the majority of the submissions are labeled \textit{Unrelated} (do not contain harmful content).
As is clear from the figure, the submission behavior of \qusers
differ in different types of subreddits, and the differences between subreddits are significant in both plots~($\chi^2(30, N=1,571)=897$, $p<0.001$ and $\chi^2(48, N=1,571)=1,086$, $p<0.001$).
In the \textit{QAnon} subreddits,
the most common type of the submissions~(34\%) are of \textit{QAnon conspiracy}.
However, 
in subreddits with \textit{Sympathetic} and those with \textit{Related (aligned)} relation labels
(the second and third group of columns from left in Figure~\ref{fig:submission}(a)),
the majority~(55\% and 65\% respectively) of the submissions by these users are of \textit{Low-quality source}~(third darkest shade).
In particular, 
these \textit{Low-quality source} content submissions make up a large portion of
\qusers' submissions in subreddits labeled with different topics as well. 
As shown in Figure~\ref{fig:submission}(b),
subreddits that are about \textit{Current events}, \textit{Pro-Trump},
and \textit{Unconstricted behavior} contain 46\%, 61\%, 67\% \textit{Low-quality source} submissions, respectively.
Interestingly, in departure from the findings of previous research~\cite{datta2019extracting,kumar2018community},
\textit{Negative interactions} constituted only a small portion of the sampled \qusers'
submissions ~(less than 10\% for all types of the subreddits).
Overall, our results indicate that \qusers
tend to spread low-quality information in the broader community, causing harm in Reddit communities at large.
This finding is especially concerning
as some of the communities they are active in are 
focused on current events, such as discussions surrounding the COVID-19 pandemic.

%% file: 07_discussion.tex
\section{Discussion and Limitations}
In this paper, 
we use the \bigban event as a source of information about early users dedicated to the QAnon conspiracy. 
We identify \bannedSubs on Reddit and 
users who were active in these communities.
We then characterize the activities of these early QAnon users beyond their these communities, from a year before the emergence of the conspiracy theory to 72 hours after the 2021 U.S. Presidential Inauguration.
Our analyses provide the first clues towards understanding whether these users are "casual conspiracists" who participate in conspiracy-related conversations occasionally and have a regular presence in other 
communities,
or "committed conspiracists" who join the platform only to participate in the activities that are related the conspiracy.
The findings point towards the latter. We discuss these findings below, and review the limitations of this work.

While not focused on the efficacy or outcomes of the ban, 
our findings echo some of the earlier results on the effect of community bans.
Previous research~\cite{chandrasekharan2017you,saleem2018aftermath} has shown that Reddit bans were effective in reducing participation from users of the banned communities and preventing some harmful behaviors. 
Our data and findings suggest that the \bigban has similar effects on the 
early \qusers community.
We show that most of 
the \qusers 
accounts we could observe
stopped being active or significantly decreased their activity after the ban.
Further, a higher percentage of accounts from \uq relative to \qauth no longer exist, implying account deletion by the most active in the \bannedSubs. One unexpected finding was the higher percentage of banned or suspended \qauth relative to \uq. In the absence of metadata or other mechanisms to understand the reasons for account suspension, future work could consider the final posts of these users to uncover whether or not the moderation intervention was related to their QAnon association.

While most researchers have tracked the activities of banned community members \textit{after}~\cite{saleem2018aftermath,jhaver2021evaluating,ali2021understanding,ribeiro2020auditing,newell2016user,rogers2020deplatforming,rauchfleisch2021deplatforming} the ban or \textit{right before} the ban~\cite{chandrasekharan2017you}, 
our analysis goes earlier in the existence of these communities to show that the majority of the \qusers likely \textit{joined} Reddit to engage in and focus on QAnon-related activities.
This finding extends our understanding of conspiracist
behavior on Reddit, and provides a recommendation for future works to study the other aspects of the participation of active members \textit{prior} to the creation of the later-banned communities, or \textit{in parallel} to their activities there.
One particular research question that this result raises is of the association between the 
success of a community ban and the degree of its participants 
engaging with other communities on the site.

Our analyses on the more-active \qusers show that while these users were involved in other communities, they concentrated their participation in subreddits with similar viewpoints.
Previous research had examined inter-community interactions and conflicts between communities, including on Reddit.
Besides some of the interactions shown in previous work, like coordinated hate attacks~\cite{mariconti2019you} and toxic conversations~\cite{kumar2018community,datta2019extracting},
we have shown that the \qusers
also engage in other types of harmful behaviors when interacting with communities that are of misaligned topics,  
such as propagating mis- and
disinformation from low-quality sources and spreading conspiracy theories. 
At the same time, despite the majority of \bannedSubs being banned for harassment (Appendix~\ref{ap:codebooks}, Table~\ref{tab:bannedSubs}), we have found only a small percentage of \qusers' submissions contain negative interactions and hate speech.
According to our results, 7\% and 0.6\% of the sampled submissions by these users
are labeled \textit{Negative interaction} and \textit{Hate speech} respectively.
Nevertheless, as \qusers made 1,554,183 submissions in the nearly 13,000 subreddits in which they were active over the 50-month study period, this 
would mean that there were over 118,000 submissions with
negative interaction
or hate speech content, 
a large amount that could still have significant impact on the broader community. 
This finding of somewhat limited proportion of negative interaction and hate speech could be a result of our sample, or the fact that our analysis only focused on Reddit  submissions, not taking comments into account. 
It is likely that there is more harassment happening in comments posted by \qusers.
A more in depth comparison of \qusers'
content, submissions and comments included,  could be an area of future work.

There are other ways in which these users can impact the overall Reddit community. As we report above, our results indicate that several of the \qusers act as moderators 
of non-QAnon subreddits. 
Content moderation plays an important role on Reddit, with moderators often having outsized influence on the content posted~\cite{juneja2020through, matiasBlackout}.
It is important for future work to understand what is the \qusers' impact on those communities and what harms to our information ecosystem may be caused by this kind of involvement creates. 

Our study focused on Reddit communities and early, committed adopters of the QAnon conspiracy theory on the platform. 
On other social media platforms, even those that offer similar "community" structures,
different platform affordances 
enable different types of user expressions~\cite{plantin2018infrastructure,van2013culture,langlois2013research}.
For example, Facebook groups also support community discussions.
However, unlike Reddit, people can be added to groups, and group activities are well integrated --- often driven by --- Facebook's existing social graph~\cite{ma2019people}. 
Such features might lead to more nuances when analyzing QAnon adherents' behaviors on Facebook~\cite{bodner2020covid}. 
Further, 
community migrations caused by deplatforming~\cite{chandrasekharan2017you,ribeiro2020auditing,ali2021understanding} raise interesting questions of how QAnon users
may exploit the affordances of various platforms. 
The activities of 
QAnon community members beyond their community boundaries in other social media platforms and across multiple platforms is another area of future work.

Nevertheless, the patterns of community and content engagement we see on Reddit might resemble that of QAnon user interactions with the broader information ecosystem in other places, online and offline.
For example, recent studies of Twitter activities around the 2020 U.S. elections~\cite{abilov2021voterfraud2020,CenterforanInformedPublic2021} have shown that on Twitter, the QAnon community forms a tight network community.
Echoing our findings, showing that Reddit QAnon users engage in large amount of the activities within sympathetic subreddits,
the QAnon community on Twitter is also quite integrated with other alt-right and extremist groups~\cite{abilov2021voterfraud2020,CenterforanInformedPublic2021}
holding sympathetic views towards QAnon.
We believe that our dataset and analyses may add knowledge to the growing research body on understanding the online alt-right activities~(e.g. \cite{papasavva2020raiders,papasavva2021qoincidence,ribeiro2020auditing,abilov2021voterfraud2020}), 
and help researchers from multiple disciplines better understand this consequential movement.

\paragraph{Limitations.} 
Our analysis suffers from several limitations worth considering here, beyond the focus on Reddit and its limited generalization to other platforms.
First, our observational data cannot be used to interpret causal relations among the observed actions. For example, \uq increased
participation in \textit{Sympathetic} subreddits post-ban relative to their pre-ban activity. Given how the data is presented, this might be interpreted as a result of the \bigban. However, there could be other explanations, such as an increase in political subreddit submissions discussing U.S. elections which also occurred during the study period.
Further, we note some limits on data collection. For example, we could not retrieve the subreddit metadata from subreddits that are currently banned, quarantined, or private. 
This limit may have impacted our data analysis in ways that are hard to estimate. 
Finally, we used the \bigban event and the \bannedSubs
to derive the set of \qusers
we built on in this work. 
There are other Reddit users who are interested in the QAnon discussions,
whose behaviors are not fully captured and analyzed in these results,
as we only consider a specific subset of communities.
An opposite phenomena may have also impacted our results: there could also be non-QAnon members who authored submissions in the subreddits being used to derive \qusers, leading us to include such accounts in the analysis. However, we estimate that the likelihood of such inclusion is low. 

%% file: 08_conclusion.tex
\section{Conclusions}
We focus our work on the consequential QAnon conspiracy theory and in particular its early, dedicated users,
those who engaged in the QAnon movement before 2018.
We characterized these users' participation on Reddit, especially outside of the \bannedSubs.
We collected a large-scale dataset of Reddit activities from identified QAnon users. 
Our results indicate that
the majority of these early QAnon users were only active on Reddit after the emergence of the QAnon conspiracy theory and decreased in activity after Reddit's \bigban.
Our qualitative analysis further shows that the more active QAnon users
engage with a wide range of subreddits focusing on various topics,
that are of different relations to the QAnon conspiracy theory.
However, 
the majority of these users' submissions were concentrated in subreddits that hold sympathetic attitudes towards QAnon, including topics that are 
supportive of Trump or unconstricted behaviors.
These submissions often include links from low-quality sources, potentially propagating mis- and disinformation.
We also found that
some of the early QAnon users become moderators in subreddits, including those that are of topics that are misaligned with the QAnon narrative.
These results improve our understandings of the QAnon community 
and their potential influence on the broader Reddit ecosystem,
providing insights for both future researchers and platform providers.

%% file: 98_acknowledgements.tex
\section{Acknowledgements}
We thank Nicola Dell for her consultation on this project. 
This material is based upon work partially supported by the National Science Foundation under grants SaTC-2120651 and IIS-1840751. 

%% file: 99_appendix.tex
\section{Data and Codebooks}

\label{ap:codebooks}
\begin{table}[H]
\footnotesize
  \caption{Selected \bannedSubs.}
  \label{tab:bannedSubs}
  \addtolength{\tabcolsep}{-2pt}    
  \begin{tabular}{lrrp{1.37cm}lrrp{1.3cm}}
    \toprule
subreddit & \begin{tabular}[c]{@{}r@{}}submission\\ authors\end{tabular} & submissions & banned for & subreddit & \begin{tabular}[c]{@{}r@{}}submission\\ authors\end{tabular} & submissions & banned for \\ 
    \midrule
\texttt{\tiny{greatawakening       }} & 12862                                                                             & 97644                            & violence harassment doxxing   & \texttt{\tiny{WWG1WGA01            }}                     & 18                                                                                & 207                              & \begin{tabular}[c]{@{}l@{}}content - \\ unspecified\end{tabular}\\
\texttt{\tiny{The\_GreatAwakening  }} & 698                & 3984                  & harassment                    &
\texttt{\tiny{TheCalmBeforeTheStorm}} & 306                & 1978                  & harassment                  \\
\texttt{\tiny{AFTERTHESTQRM        }} & 46                 & 745                   & harassment                    &
\texttt{\tiny{BiblicalQ            }} & 142                & 464                   & harassment                    \\
\texttt{\tiny{TheGreatAwakening    }} & 12                 & 391                   & harassment                    &
\texttt{\tiny{thestorm             }} & 33                 & 252                   & ban \\
 & & & &
 & & & evasion                   \\
\texttt{\tiny{QAnon                }} & 59                 & 122                   & harassment                    &
\texttt{\tiny{QProofs              }} & 2                  & 79                    & harassment                    \\
\texttt{\tiny{QGreatAwakening      }} & 23                 & 34                    & spam                          &
\texttt{\tiny{QanonandJFKjr        }} & 14                 & 31                    & spam                          \\
\texttt{\tiny{QanonUK              }} & 12                 & 27                    & harassment                    &
\texttt{\tiny{qresearch            }} & 15                 & 19                    & harassment                    \\
\texttt{\tiny{QanonTools           }} & 3                  & 18                    & harassment                    &
\texttt{\tiny{NorthernAwakening    }} & 5                  & 13                    & harassment                    \\
\texttt{\tiny{greatawakening2      }} & 6                  & 7                     & ban &
\texttt{\tiny{Q4U                  }} & 2                  & 5                     & ban \\
 & & & evasion                   &
 & & & evasion                   \\
\texttt{\tiny{Quincels             }} & 4                  & 5                     & ban \\
& & & evasion \\
\bottomrule
\end{tabular}
  \addtolength{\tabcolsep}{1pt}    
\end{table}

\begin{table}[H]
\caption{\label{tab:subCodes} Codebook for subreddits.}
\footnotesize
\begin{tabular}{lp{0.37\textwidth}l}
\toprule
Code &                                                                                         Code Description & Example Subreddits \\
\midrule
anti-establishment            &                                             Against the norms, against the institutions, anti-governance & 4chan, wallstreetbets \\
anti-hate speech              &                                                     Empowering those targeted by hate speech & lgbt, TwoXChromosomes \\
anti-left                     & Against left leaning politics or politicians, against progressive concepts or figures/organizations                                                                                                   & Liberalstupidity, DNCLawsuit \\
anti-right                    &    Against right leaning politics or politicians, against conservative concepts or figures/organizations                                                                                                     & FoxFiction, Republicancer \\
anti-Trump                    &  Against the 45th US President                                                                                                       & LyingTrump, MarchAgainstTrump \\
Christianity                  &  Discuss Christianity or are defined by a membership of current/former believers                                                                                                 & Christianity, ChristiansAwake2NWO \\
conspiracy - other            &     Devoted to, discussing, or promoting conspiracy theories not related to QAnon or other political conspiracy                                                                                                   & aliens, ConspiracyZone \\
conspiracy - politics         &     Devoted to, discussing, or promoting politics-related conspiracy theories not related to QAnon                                                                                                                           & 911truth, PoliticalConspiracy \\
coronavirus                   &  Discussing or providing updates about the COVID-19                                                                                                       & Coronavirus, China\_Flu \\
apocalypse                    &                        Crumbling society, dark age, Armageddon, apocalyptic, end times, new world order 
& Apocalypse, collapse \\
hate speech                   &  Anti-Arabic, anti-diversity, anti-Semitic, anti-globalisation, pro-nationalist, sexism, white supremacy 
& MGTOW, CringeAnarchy \\
metaphysical/wellness         &                   Beyond perception, open to the unseen  
& PsychedelicMessages, zen \\
news                          &   Sharing articles and stories about current events                                                                                                      & news, CorruptioNNews \\
politics - left               &   Supporting ideas, a political movement or party, and politicians from the center-left to far-left on the US political spectrum                                                                                                      & democrats, WayOfTheBern \\
politics - other              &         Political movement or party, third party or international, not related to US left or right sides & ukpolitics, Israel \\
politics - right              &     Supporting ideas, a political movement or party, and politicians from the center-right to far-right on the US political spectrum                                                                                                         & Republican, new\_right \\
politics - unaffiliated       &                                                              About politics, no strong direction leaning & politics, The\_Mueller \\
pro- high quality information &                                                                     Pro-moderation, high-quality sources & Coronavirus, Qult\_Headquarters \\
pro-freedom of speech         &                                                                     Anti-moderation, low-quality sources & LibertarianUncensored, 4chan \\
pro-Trump                     &   In support of the 45th US President, identifying as Trump supporters                                                                                                      & The\_Donald, DrainTheSwamp \\
QAnon                         &   Q, the Great Awakening, and other elements central to the QAnon narrative                                                                                                      & CBTS\_Stream, QRedPills \\
save the children             &                                                                            Pedophilia, human trafficking & PedoGate, adrenochrome \\
the unexplained               &                                                                                   Mysteries, the unknown & theunexplained, CoincidenceTheorist \\
unrelated                     &                                                                                                 Does not fit into existing labels
& xboxone, vegan \\
\bottomrule
\end{tabular}
\end{table}

\begin{table}[H]
\caption{\label{tab:subRelation} Subreddit code to relation labels map.}
\footnotesize  
\begin{tabular}{p{0.2\textwidth}p{0.45\textwidth}p{0.25\textwidth}}
\toprule
Relation Label             & Label Description                                                                                          & Codes                 \\
\midrule
QAnon              & QAnon narrative                                                                                     & QAnon, save the children\\                         
Sympathetic        & Having similar views or support common QAnon themes.                                                 & conspiracy - other, conspiracy - politics, anti-establishment, anti-left, politics - right, pro-freedom of speech, pro-Trump         \\
Related (aligned)    & Share similar elements that are directly aligned with QAnon views or common themes                  & apocalypse, metaphysical/wellness, Christianity, the unexplained, hate speech                   \\
Related (neutral)    & Sharing similar elements, but not evidently aligned or misaligned with QAnon views or common themes & coronavirus, new, politics-other, politics-unaffiliated       \\
Related (misaligned) & Sharing similar elements, but directly misaligned with QAnon views or common themes                                                                                                    & anti-right, anti-Trump, politics-left, pro-high quality information, anti-hate speech              \\
Unrelated                           & Does not fit into existing relationships                                                                              & unrelated  \\ 
\bottomrule
\end{tabular}
\end{table}

\begin{table}[H]
\caption{\label{tab:subTopics} Subreddit code to topic labels.}
\footnotesize  
\begin{tabular}{p{0.2\textwidth}p{0.45\textwidth}p{0.25\textwidth}}
\toprule
Topic Label                                         & Label Description                                                                                                                                                                     & Codes                          \\
\midrule
QAnon & Mentions of Q, pedophilia, human trafficking, cannibalism, satanic ritual                                                                                            & QAnon, save the children             \\
Unseen/unknown              & Speculation or searching for meaning or explanation about: Christianity, metaphysical/wellness, the unexplained, conspiracies. & conspiracy - other,  conspiracy-politics, Christianity, metaphysical/wellness, apocalypse, the unexplained               \\
Pro-Trump        & Politics with a history of supporting Trump & pro-Trump, politics-right, anti-left                     \\
Trump (unknown) &  Politics where Trump affiliation is unknown or not relevant                                                                                                                                                                    & politics - other, politics-unaffiliated       \\
Anti-Trump          & Politics with a history of opposing Trump                                                                                                                            & politics - left, anti-right, anti-Trump                    \\
Current events                     & News without a stance on speech/moderation, coronavirus.                                                                                                             & news, coronavirus                   \\
Constricted behavior               & Pro-high-quality information / pro-high quality information / moderated news                                                                                        & anti-hate speech, pro-high quality information \\
Unconstricted behavior             & Pro-freedom of speech / free speech over information quality / hate speech / unmoderated news (communication based)                                                  & hate speech, pro-freedom of speech, anti-establishment\\
Unrelated                                           & Unaffiliated with other topics                                                                                                                                                        & unrelated   \\
\bottomrule
\end{tabular}
\end{table}

\begin{table}[H]
\footnotesize
\caption{\label{tab:submission_code} Codebook for submissions.} 
\begin{tabular}{lp{0.37\textwidth}p{0.37\textwidth}}
\toprule
Harmful Content Label           & Label Description  & Example Submission\\
\midrule
QAnon conspiracy               &       Q, the Great Awakening, and other elements central to the QAnon narrative                                                                                                                                                                                                                                                                                                                                                                                                                                                                                                                                                                           & ``Latest Qanon-Related Developments 7-24-18 {stroppy me}'' \\
Hate speech          & 
Derogatory or discriminatory language against identity-based attributes such as color, disability, ethnicity, gender identity, nationality, race, religion, sex, and sexual orientation
 & ``What the Bible Really Says About Homos'' \\
Conspiracy-politics &  Devoted to, discussing, or promoting politics-related conspiracy theories not related to QAnon                                                                                                                                                                                                                                                                                                                                                                                                                                                                                                                                                                                & ``212.5. Hillary's Leakers and Hackers, Awan Brothers Saga Deepens'' \\
Conspiracy-other     &     Devoted to, discussing, or promoting conspiracy theories not related to QAnon or other political conspiracy                                                                                                                                                                                                                                                                                                                                                                                                                                                                                                                                                                              & ``Interesting interview on the Vatican infiltration. I can't vouch for the trustworthiness of the people (Westall, Rothstein), but maybe you can separate the wheat from the chaff'' \\
Low-quality source   & Labeled by Media Bias Fact Check as ``Low Credibility'', ``Mixed Factual Reporting'' and below, Questionable Sources, or Conspiracy-Psuedoscience                                                                                                                                                                                                                                                                                                                                                                                                                                    & ``Jerry Nadler Subpoenas Corey Lewandowski for Testimony'', linked article from Breitbart \\
Negative interaction & Accusations or attack on a person or group based on views or non-protected characteristics affiliation                                                                                                                                                                                                                                                                                                                                                                                                                                                                          & ``Swalwell Inexplicably Wades Back Into His Own Diarrhea Pit on His Nuke Threat'' \\
Unrelated            &  Unaffiliated with other harm labels                            & ``Mercedes-Benz apologizes to Chinese for quoting Dalai Lama'' \\              \bottomrule                                              
\end{tabular}
\end{table}